\documentclass[aps,pre,twocolumn,showpacs,preprintnumbers,superscriptaddress,amsmath,amssymb]{revtex4-1}

\usepackage[T1]{fontenc}
\usepackage[latin1]{inputenc}
\usepackage{graphicx}
\usepackage{amsfonts, amsmath}
\usepackage{amssymb,mathtools}
\usepackage{amstext}
\usepackage{color}
\usepackage{dsfont}
\usepackage{subfig}

\newcounter{commentzaehler}
\begin{document}
\graphicspath{{figs/}}
\preprint{SAW16, dated \today}

\title{Synchronization of organ pipes}

\author{Jakub Sawicki}
\email[corresponding author: ]{zergon@gmx.net}
\affiliation{Institut f{\"u}r Theoretische Physik, Technische 
Universit{\"a}t Berlin, Hardenbergstra{\ss}e 36, 10623 Berlin, 
Germany}
\author{Markus Abel}
\affiliation{Department of Physics and Astronomy, Potsdam University, Karl-Liebknecht-Stra{\ss}e 24, 14476 Potsdam 
Germany}
\author{Eckehard Sch\"{o}ll}
\affiliation{Institut f{\"u}r Theoretische Physik, Technische
  Universit{\"a}t Berlin, Hardenbergstra{\ss}e 36, 10623 Berlin, Germany}


% Synchronization, nonlinear dynamics, 05.45.Xt
% neural networks, 87.85.dq
% Delay equations, in function theory, 02.30.Ks
% Complex systems, 89.75.-k

\begin{abstract} 
We investigate synchronization of coupled organ pipes. Synchronization and reflection in the organ lead to undesired weakening of the sound in special cases. Recent experiments have shown that sound interaction is highly complex and nonlinear, however, we show that two delay-coupled Van-der-Pol oscillators appear to be a good model for the occurring dynamical phenomena. Here the coupling is realized as distance-dependent, or time-delayed, equivalently. Analytically, we investigate the synchronization frequency and bifurcation scenarios which occur at the boundaries of the Arnold tongues. We successfully compare our results to experimental data.
\end{abstract}

\maketitle

{\em Introduction -} The physics of organ pipes is an interdisciplinary topic where many fields of science meet. It is highly interesting as it includes elements of nonlinear dynamical system theory \cite{FAB00,BAD13,FLU13}, aeroacoustic modeling \cite{HOW03} and synchronization theory \cite{PIK01}. The focus of these different research areas is the ``queen of instruments'' which captivates through the grandeur of her sight and majesty of her sound. Here, we investigate an interesting nonlinear effect: organ pipes close to each other synchronize. Recent studies have been of experimental nature as well as theoretical \cite{ABE06,ABE09,FIS14,FIS16}. For musical purposes, synchronization of sound might be desired or not: it might stabilize the pitch of special organ pipes as a favorable effect, whereas sound weakening, as observed in the prospect of an organ, is highly undesired \cite{RAY82,FLE78,STA01}. This weakening occurs as an amplitude minimum due to destructive interaction between pipes during the actuating of the swell box, where the pipes stand close to each other. 

A qualitative understanding of the nonlinear mechanisms is obtained following the arguments of \cite{ABE09,FIS14}: a single organ pipe can be described as a self-sustained oscillator, where the oscillating unit consists of the jet, or ``air sheet'' which exits at the pipe mouth. The resonator is of course the pipe body; there, sound waves emitted at the labium (i.e., the sharp edge in the upper part of the pipe's opening) travel up and down and can trigger a regular oscillation of the air sheet. Energy is supplied by the generating unit, which is the pressure reservoir beneath the pipe at a basically constant rate. Experimental and numerical investigations by Abel et al. \cite{ABE09} yield the conclusion that an organ pipe can be approximated satisfactorily by a Van der Pol oscillator. Whereas the work of Fischer \cite{FIS14} focuses on the nonlinearities in sound generation and their effect on the synchronization properties, in this paper we investigate the effect of the finite distance of two coupled pipes which in turn is reflected by a delay in the coupling function. More specifically, we investigate the bifurcation scenarios in the context of two delay-coupled Van-der-Pol oscillators as a representation of the system of two coupled organ pipes, such as in the experimental setup of Bergweiler et al. \cite{BER06}. In extension of previous work, we study the dependence of Arnold tongues under variation of the time delay $\tau$ and the coupling strength $\kappa$, to explore how undesired synchronization or chaotic behavior can be avoided. We compare our results to experimental measurements of the synchronization under variation of the pipe distance and find a qualitative coincidence of the nonmonotonic modulation of the shape of the Arnold tongue which is in contrast to the linear boundaries of the Arnold tongues for systems with undelayed coupling. 

In Section II we introduce two delay-coupled Van der Pol oscillators as a simple model of coupled organ pipes. In section~\ref{Analytic approaches} we apply two analytical methods to get a better understanding of the synchronization phenomena. The central part of the paper is Sec.~\ref{sec: synchronization}, where we present the analytical results. In Sec.~\ref{sec:comparison} we compare these results with acoustic experiments. We conclude with Sec.~\ref{sec:conclusion}.

%%%%%%%%%%%%%%%%%%%%%%%%%%%%%%%%%%%%%%%%%%%%%%%%%%%%%%%%%%%%%%%%%%%%%%
\section{A model of coupled organ pipes}
\begin{figure}%[H]
%\centering
\includegraphics[width=1.04\linewidth]{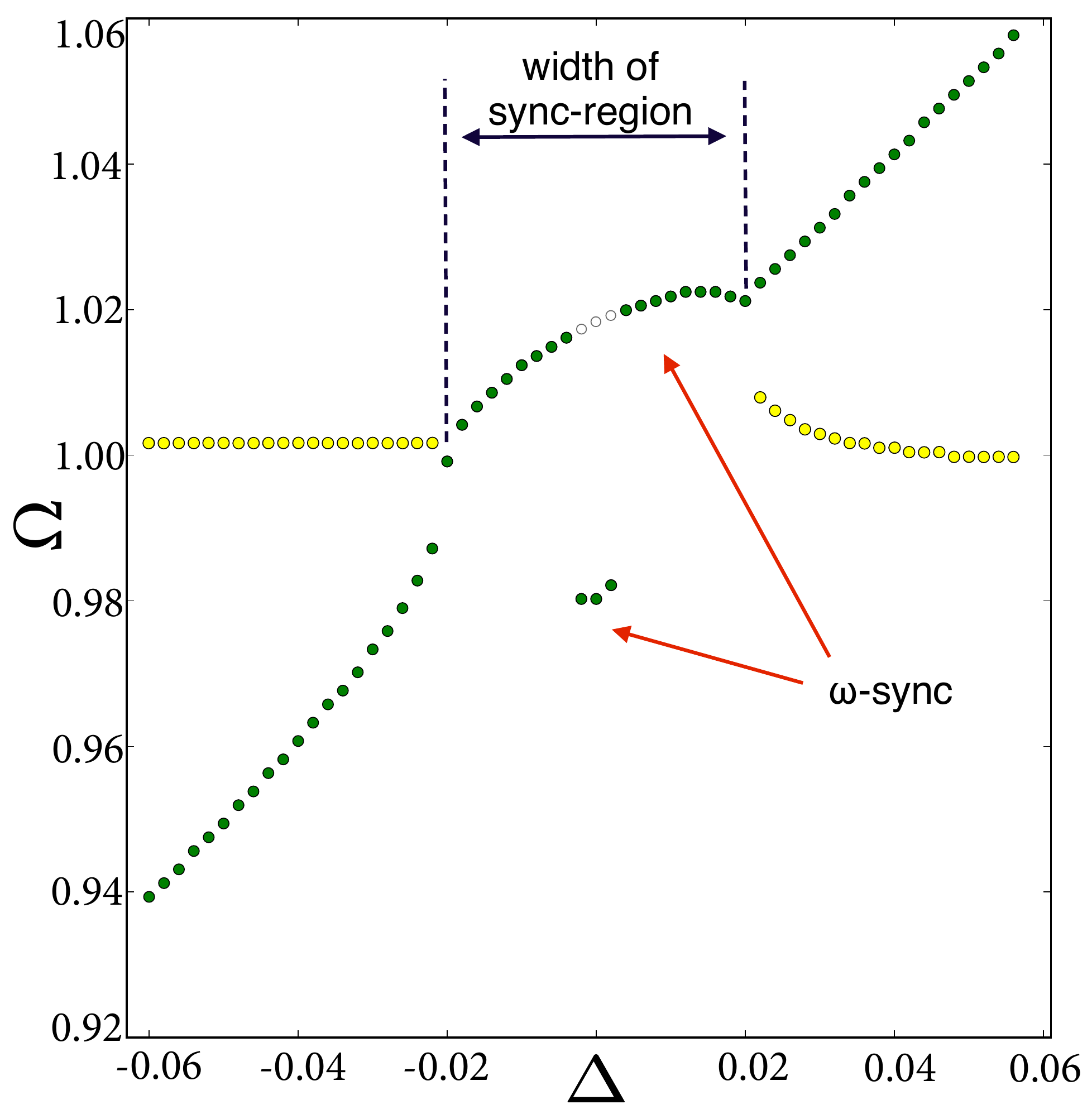}
\caption{\label{goal}\raggedright Angular frequency $\Omega$ of oscillator $x_1$ (dark green circles), and oscillator $x_2$ (light yellow circles) versus the detuning $\Delta$ of the oscillators. Full (empty) circles correspond to symmetric (non-symmetric) initial conditions. Parameters:  $\omega_{2}=1$, $\mu = 0.1$, $\gamma=1$, $\kappa=0.4$, $\tau=0.1\pi$. 
}
\end{figure}

To obtain a deeper insight into the synchronization phenomena of two coupled organ pipes we model the pipes by Van der Pol oscillators with delayed cross-coupling:
\begin{equation}
\label{ausgang}
\ddot{x}_i+{\omega_i}^2x_i-{\mu}\left[\dot{x}_i-\dot{f}(x_i)+{\kappa}x_j(t-\tau)\right]=0,
\end{equation}
where $i,j=1,2$. These equations represent a harmonic oscillator with an intrinsic angular frequency $\omega_i$, supplemented with linear and nonlinear damping of strength $\mu > 0$. The nonlinear damping can be described by the nonlinear function
\begin{eqnarray}
\label{ausgang11}
f(x_i)=\frac{{\gamma}}{3}x_i^3,
\end{eqnarray}
where $\gamma$ is the anisochronicity parameter and $\dot{f}(x_i)={\gamma} x_i^2 \dot{x_i}$. The coupling strength in Eq.\,\eqref{ausgang} is $\kappa$, and the coupling delay is $\tau$. Since for synchronization the frequency difference of the two oscillators is important, we introduce the detuning parameter $\Delta \in \mathbb{R}$ by
\begin{eqnarray}
\label{ausgang12}
\omega_1^2=\omega_2^2+\mu \Delta. 
\end{eqnarray}
\\
In Fig.\,\ref{goal} we display frequency locking as obtained by numerical simulation of Eq.\,\eqref{ausgang} with symmetric initial conditions . The observed angular frequencies $\Omega$ are plotted versus the detuning ${\Delta}$ of two Van der Pol oscillators.  
A pronounced synchronization region and a sharp transition to synchronization is observed. Within the synchronization region, only for small $|\Delta|$ the in-phase synchronized solution is observed (lower branch), while for larger $|\Delta|$ the anti-phase synchronized solution (upper branch) is obtained even for symmetric initial conditions (full circles). Note that for small $|\Delta|$ the anti-phase synchronized solution is also observed for any non-symmetric initial conditions (empty circles).
Our purpose is to analyze the synchronization frequency, the width of the synchronization region, the phase difference in the synchronized state, its stability, and the bifurcation scenarios which occur at the boundaries.

%%%%%%%%%%%%%%%%%%%%%%%%%%%%%%%%%%%%%%%%%%%%%%%%%%%%%%%%%%%%%%%%%%%%%%%%%%%%%%%%%%%%%%%%%%%%%%%%

\section{Analytic approaches}
\label{Analytic approaches}
\subsection{Method of averaging}

%\todo{(13): Bei Laplace Trafo die Anfangsbedingung 0 setze. Darf ich das?}

The method of averaging (quasiharmonic reduction) describes weakly nonlinear oscillations in terms of slowly varying amplitude and phase.\\
For $ {\mu}= 0$ the uncoupled system reduces to the harmonic oscillator $\ddot{x}_i+{\omega_i}^2x_i=0$ with solution
\begin{eqnarray}
\label{hl}
{x_i}&=&R_i\sin({\omega_i}t+\phi_i),
\end{eqnarray}
with constant amplitude $R_i$ and phase $\phi_i$. 
For $0<{\mu}\ll 1$ we look for a solution in the form Eq.~(\ref{hl}) but assume that the amplitude $R_i \ge 0$ and the phase $\phi_i$ are time-dependent functions:

\begin{equation}
\label{fhl1}
\begin{split}
{x_i}=R_i(t)\sin({\omega_i}t+\phi_i(t)),\quad\\
\dot{x}_i=R_i(t){\omega_i}\cos({\omega_i}t+\phi_i(t)).
\end{split}
\end{equation}
where terms involving the slowly varying functions $\dot{R_i}$, $\dot{\Phi_i}$ are neglected.
Without loss of generality, we choose ${\omega_{2}} = 1$. For small $\mu$ we use the method of averaging, assuming that the product $\mu\tau$ is small, and Taylor expand $R_i(t-\tau)$ and $\phi_i(t-\tau)$ in the following way:
\begin{eqnarray}
\label{taylor1}
R_i(t-\tau)=R_i(t)-\tau\dot{R}_i(t)+\frac{\tau^2}{2}\ddot{R}_i(t)+\dots,
\end{eqnarray}
We introduce the phase difference $\psi(t)=\phi_1(t)-\phi_2(t)$. Defining a new time scale $\tilde{t}=\frac{2t}{\mu}$ we find the equations which describe the system \eqref{ausgang} on a slow time scale:
\begin{eqnarray}
\label{R1}
\dot{R}_{1/2}(\tilde{t})=&{R_{1/2}(\tilde{t})}\left(1-\frac{{\gamma}{R_{1/2}(\tilde{t})}^2}{4}\right) \nonumber \\&\mp{\kappa}R_{2/1}(\tilde{t})\sin(\psi(\tilde{t})+\tau),
\end{eqnarray}
\begin{eqnarray}
\label{Psi}
\dot{\psi}(\tilde{t})=&-\Delta+{\kappa}\left[\frac{R_1(\tilde{t})}{R_2(\tilde{t})}\cos(\psi(\tilde{t})-\tau)\right. \nonumber \\ &\left.-\frac{R_2(\tilde{t})}{R_1(\tilde{t})}\cos(\psi(\tilde{t})+\tau)\right].
\end{eqnarray}
For simplicity, in the following we omit the tilda\,$\sim$. The method of averaging together with truncation of the Taylor expansion in $\tau$ reduces the infinite-dimensional problem to a finite-dimensional problem by assuming that the product $\mu\tau$ is small. This key step enables us to handle the original delay differential equation as a system of ordinary differential equations \cite{WIR02,SEM15}. We now have two dynamical equations \eqref{R1} for the amplitudes $R_1$ and $R_2$, and one equation \eqref{Psi} for the phase difference $\psi(t)$ which is also called {\em slow phase}. The latter equation is a generalized Adler equation \cite{ADL73} and contains the main features of synchronization.\\

\begin{figure}
%\centering
\includegraphics[width=1.0\linewidth]{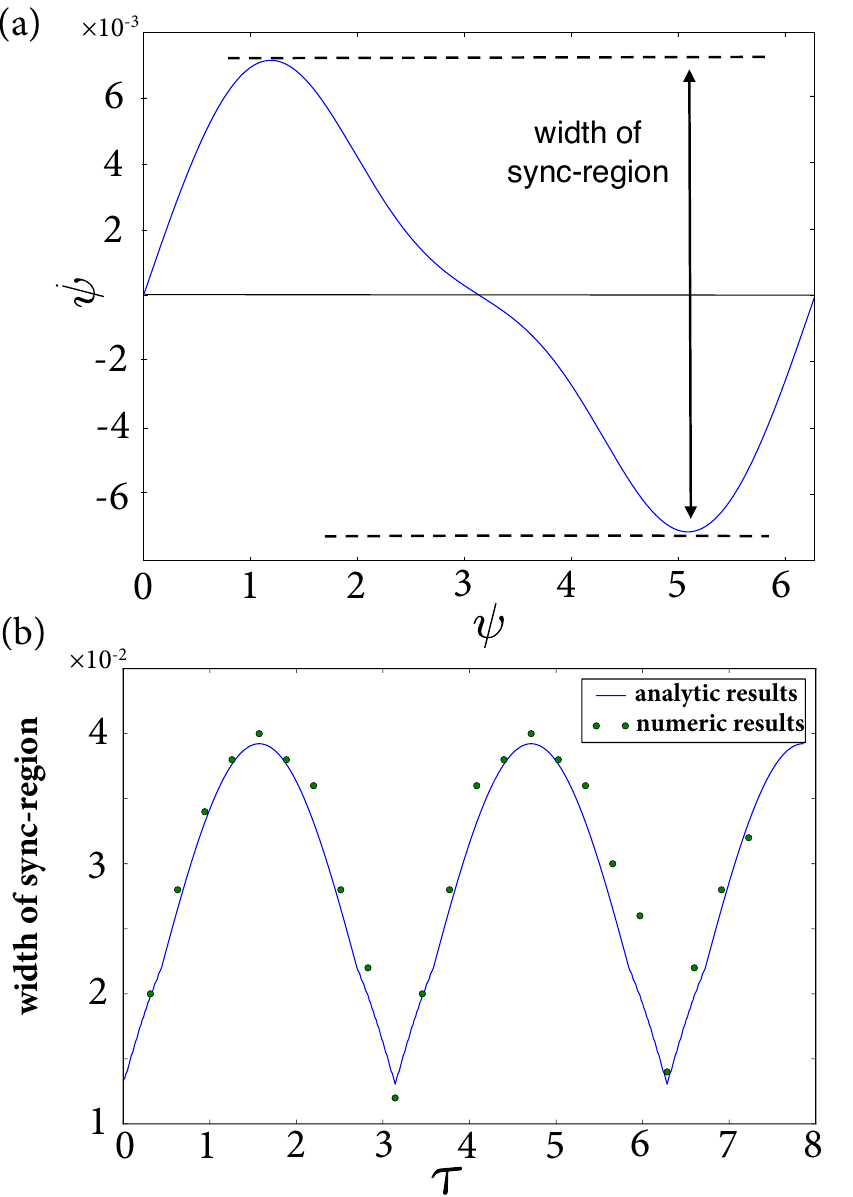}
\caption{\label{adlerplot2}\raggedright (a) The right-hand side of the Adler equation~\eqref{adler} $\dot{\psi}=q({\psi})$ for zero frequency detuning $\Delta=0$ versus the slow phase $\psi$. The difference between the maximum and minimum gives the width of the synchronization tongue. Parameters: $\omega_{1}=\omega_{2}=1$, $\mu = 0.1$, $\gamma=1$, $\kappa=0.2$, $\tau=0.1\pi$. (b) Analytic (line) and numeric (dots) results for the width of the synchronization tongue as a function of the delay time $\tau$ for $\omega_{2}=1$, $\mu = 0.1$, $\gamma=1$, $\kappa=0.4$.}
\end{figure}

{\em Generalized Adler equation -}
The equilibria of the Adler equation correspond to the locking of phase and frequency, since the difference between the phases is constant. To investigate the stability and bifurcation scenario of such fixed points we take a closer look at the generalized Adler equation\eqref{Psi}, written in general form:
\begin{equation}
\label{adler}
\begin{split}
\dot{\psi}(t)=-\Delta+{\kappa}q({\psi}(t))
\end{split}
\end{equation}
where the averaged forcing term $q({\psi}(t))$ is the $2\pi$-periodic function
\begin{equation}
\label{q}
\begin{split}
q({\psi}(t))=\tfrac{R_1(t)}{R_2(t)}\cos\left[\psi(t)-\tau\right]-\tfrac{R_2(t)}{R_1(t)}\cos\left[\psi(t)+\tau\right].
\end{split}
\end{equation}
The generalized Adler equation~\eqref{Psi} is necessary for the calculation of the Arnold tongue (see below), which is one of the main characteristics of synchronization in nonlinear systems. For further analysis it is useful to eliminate the amplitudes $R_i(t)$ from Eq.\,\eqref{q}. Therefore we express $R_2(t)$ by $R_1(t)$ in the case of a relative equilibrium of the amplitudes ($\dot{R}_i(t)=0$). We achieve two relevant solutions for the stationary amplitude. Inserting the values of $R_i$ into Eq.\,\eqref{q} -- according to the numerical results -- we can plot $\dot{\psi}$ versus $\psi$ in Fig.\,\ref{adlerplot2}a, which gives a graph of the right-hand side of the generalized Adler equation\eqref{adler}, i.e., the function $q({\psi})$ in the case $\Delta=0$. 
Since in the synchronization region $\dot{\psi}=0$ and hence $\Delta=q({\psi})$, the maximum and minimum in Fig.\,\ref{adlerplot2}a correspond to the border of the synchronization tongue when varying $\Delta$ as we can see in Fig.\,\ref{goal}.  A change of $\Delta$ shifts, according to Eq.\,\eqref{adler}, the curve in the $y$-direction but does not change its shape. In this way we can calculate the width of the synchronization tongue as a function of $\tau$ and compare these analytic results to our numerical ones from the simulation of  Eqs.(\ref{ausgang}) (see Fig.\,\ref{adlerplot2}b). The agreement between the results is remarkable, even though there is an unavoidable small deviation because of the limited numerical accuracy of our simulation. The transient times are very large at bifurcation points. Furthermore we gain information about the stability of the synchronization state from Fig.\,\ref{adlerplot2}a: For $\psi=0$ (or, equivalently, $\psi=2\pi$) we have an unstable equilibrium since, and for $\psi=\pi$ (anti-phase oscillation) a stable equilibrium since $\dot{\psi}<0$ for $\psi>0$ and $\dot{\psi}>0$  for $\psi<0$, in accordance with experimental results~\cite{FIS14}, as discussed in Fig.\,\ref{realpipeex}a below. The experimentally observed decrease of the amplitude at  $\Delta = 0$  indicates an anti-phase oscillation~\cite{ABE06}.

\subsection{Describing function method}

\begin{figure}%[H]
%\centering
\includegraphics[width=.8\linewidth]{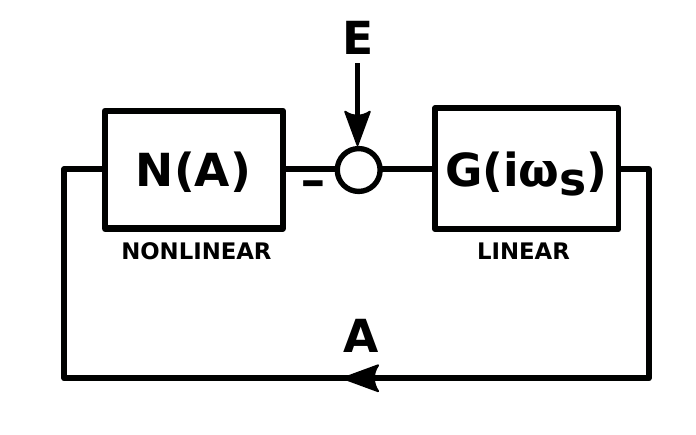}
\caption{\label{block}\raggedright Block diagram of a nonlinear feedback control loop consisting of a linear transfer function $G(i\omega_s)$ and a nonlinear describing function $N(A)$. $A$ stands for the feedback signal and $E$ for the reference input, which is added at the summation point (circle).}
\end{figure}

The describing function method (also called the method of harmonic balance) uses frequency domain techniques to investigate limit cycle behavior in nonlinear systems, which is typically represented by a block diagram (Fig.\,\ref{block}). This engineering method requires an approximation, but nevertheless often gives a reliable prediction of the frequency $\omega_s$ and amplitude $A$ of the limit cycle in a nonlinear system.\\
The condition for the harmonic balance, i.e., the limit cycle, of our system can be formulated in the following way \cite{FOL93,GHO13}:
\begin{equation}
\label{Nyq}
\underbrace{G_i^{-1}(i\omega_s)}_{LINEAR}+\underbrace{N(A_i)}_{NONLINEAR}=\frac{E_i}{A_i}e^{-i\Delta \phi_i},\quad i=1,2.
\end{equation}
$G_i(i\omega_s)$ is called the transfer function of the $i^{th}$ node and $N(A_i)$ the describing function of the nonlinear control loop, where $\omega_s$ is the angular frequency of the limit cycle and $A_i$ its amplitude. $\Delta \phi_i$ is the phase difference between the oscillation of the limit cycle and the reference input, which has the amplitude $E_i$. Equation \eqref{Nyq} holds for a static, odd 
nonlinearity and a transfer function which behaves like a low-pass filter. For the method of harmonic balance we use Eqs.\,\eqref{ausgang} and \eqref{ausgang11}:
\begin{equation}
\label{hb1}
\underbrace{\ddot{x}_i+\omega_i^2x_i-{\mu}(\dot{x}_i}_{LINEAR}-\underbrace{\dot{f}(x_i)}_{\mathclap{NONLINEAR}})={\mu}\kappa x_j(t-\tau).
\end{equation}
The Laplace transform ($X_i(i\omega_s)=\mathcal{L}\left\{x_i(t)\right\}=\int_0^{\infty}x_i(t)e^{-i{\omega_s}t}\,dt$) of Eq.\,\eqref{hb1} leads to
\begin{equation}
\label{hb3}
\underbrace{\frac{\omega_i^2-i\omega_s \mu-\omega_s^2}{i\omega_s{\mu}}}_{G_i^{-1}(i\omega_s)}{X_i}+F(X_i)=-i e^{-i\omega_s \tau} \underbrace{\frac{\kappa}{\omega_s} X_j}_{E_i},
\end{equation}
where $F(X_i)$ is the Laplace transform of the nonlinear function $f(x_i)$. The transfer function $G(i\omega_s)$ is the characteristic of the linear part of our system and can be read from Eq.\,\eqref{hb3}:
\begin{equation}
\label{transfer}
G_i(i\omega_s)=\frac{i\omega_s{\mu}}{\omega_i^2-i\omega_s \mu-\omega_s^2}.
\end{equation} 
The $\omega_s^2$-term in the denominator guarantees the low-pass behavior.
The describing function $N(A)$ is the amplification of the fundamental harmonics of the periodic signal $x_i(t)=A_i\sin(\omega_s t+\phi_i)$ by the nonlinear function $f(x_i)=\frac{{\gamma}}{3}x_i^3$ of our system:
\begin{equation}
\label{describing}
N(A_i)=\frac{F(X_i)}{X_i}=\frac{f(x_i)}{x_i}\approx\frac{{\gamma}}{4}A_i^2.
\end{equation} 
The approximation of the describing function method is based on the exclusive consideration of the fundamental harmonics in Eq.\,\eqref{describing}. For its calculation we have used the following trigonometric addition theorem: $\sin^3(\zeta)=\frac{1}{4}\left[3\sin(\zeta)-\sin(3\zeta) \right]$. According to the right-hand side of Eq.\,\eqref{hb3} the reference input of one system is given by the delayed output of the other system. The factor $-i e^{-i\omega_s \tau}$ means a negative phase shift of $\frac{\pi}{2}+\omega_s \tau$ in the time domain, so that the reference input in case of synchronization is given by $\frac{\kappa A_j}{\omega_s}\sin(\omega_s (t-\tau)+\phi_j-\tfrac{\pi}{2})$. In the case of the $i^{th}$ oscillator the right-hand side of Eq.\,\eqref{Nyq} yields
\begin{equation}
\label{reference}
\frac{E_i}{A_i}e^{-i\Delta \phi_i}=\frac{\kappa A_j}{\omega_s A_i}e^{-i\left[\phi_i-\phi_j+\omega_s\tau+\frac{\pi}{2}\right]} .
\end{equation} 
By applying Eqs.\,(\ref{transfer})-(\ref{reference}) to Eq.\,\eqref{Nyq} we obtain
\begin{equation}
\label{ws1}
\frac{i}{\mu}(\omega_s-\frac{\omega_i^2}{\omega_s})-1+\frac{{\gamma}}{4}A_i^2= \frac{\kappa A_j}{\omega_s A_i }e^{-i\left[\phi_i-\phi_j+\omega_s\tau+\frac{\pi}{2}\right]}.
\end{equation}
The imaginary part of Eq.\,\eqref{ws1} gives us information about the synchronization frequency $\omega_s$ versus the time delay. Multiplying it by the imaginary part of the analog equation for the $j^{th}$ oscillator, we obtain:
\begin{equation}
\label{freql}
\left(\omega_s-\frac{\omega_i^2}{\omega_s}\right)\left(\omega_s-\frac{\omega_j^2}{\omega_s}\right)= \frac{\mu^2\kappa^2}{\omega_s^2}\cos(\psi+\omega_s\tau)\cos(\psi-\omega_s\tau),
\end{equation}
where we have introduced the phase difference $\psi=\phi_1-\phi_2$.\\

The two analytic approaches in this section yield Eqs.\,\eqref{adler} and \eqref{freql}, respectively, whereby we can get information about the phase difference and the frequency of the coupled oscillators, respectively. We use these results in the following section.
%%%%%%%%%%%%%%%%%%%%%%%%%%%%%%%%%%%%%%%%%%%%%%%%%%%%%%%%%%%%%%%%%%%%%%%%%%%%%%%%%%%%%%%%%%%%

\begin{figure}
%\centering
\includegraphics[width=.94\linewidth]{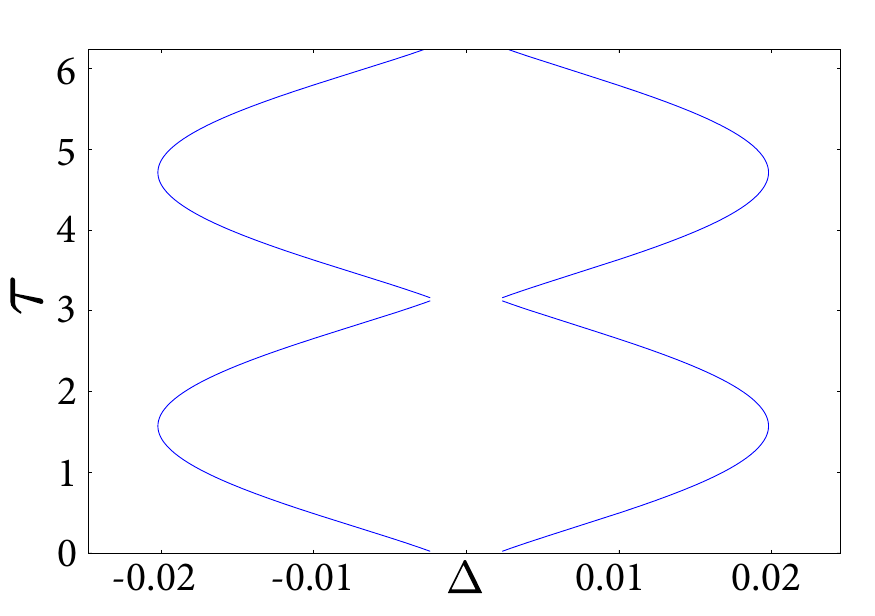}
\caption{\label{arnoldtau}\raggedright The synchronization region in the parameter plane of delay time $\tau$ and frequency detuning $\Delta$. The plot shows an analytically computed Arnold tongue for the coupling strength $\kappa=0.2$, $\omega_2=1$, $\mu = 0.1$, $\gamma=1$. One can see the symmetric, $\pi$-periodic boundaries of the tongue.}
\end{figure}

\section{Synchronization phenomena} \label{sec: synchronization}
{\em Arnold tongue -}
One important theoretical question is the transition to synchronization, usually characterized in the parameter plane of frequency detuning $\Delta$ and coupling strength $\kappa$. In the case of delayed coupling the time delay $\tau$ has an importance comparable to the coupling strength $\kappa$. The synchronization region in the $(\kappa$, $\Delta$) or ($\tau$, $\Delta$) plane is generally called Arnold tongue, and it is one of the main characteristics of synchronizing nonlinear systems. First, we keep the coupling strength $\kappa \neq 0$ constant and calculate the synchronization region analytically using the methods from the previous Section in the plane of the delay time $\tau$ and the detuning $\Delta$ (Fig.~\ref{arnoldtau}), in excellent agreement with the numeric results from Eq.~(\ref{ausgang}). The boundaries of the Arnold tongue are modulated periodically with a period $\pi$ as $\tau$ is varied. Note that for our choice of $\omega_2=1$ the period of the uncoupled harmonic oscillator is $2\pi$.
\\ 

%%%%%%%%%%%%%%%%%%%%%%%%%

\begin{figure}
%\centering
\includegraphics[width=1.02\linewidth]{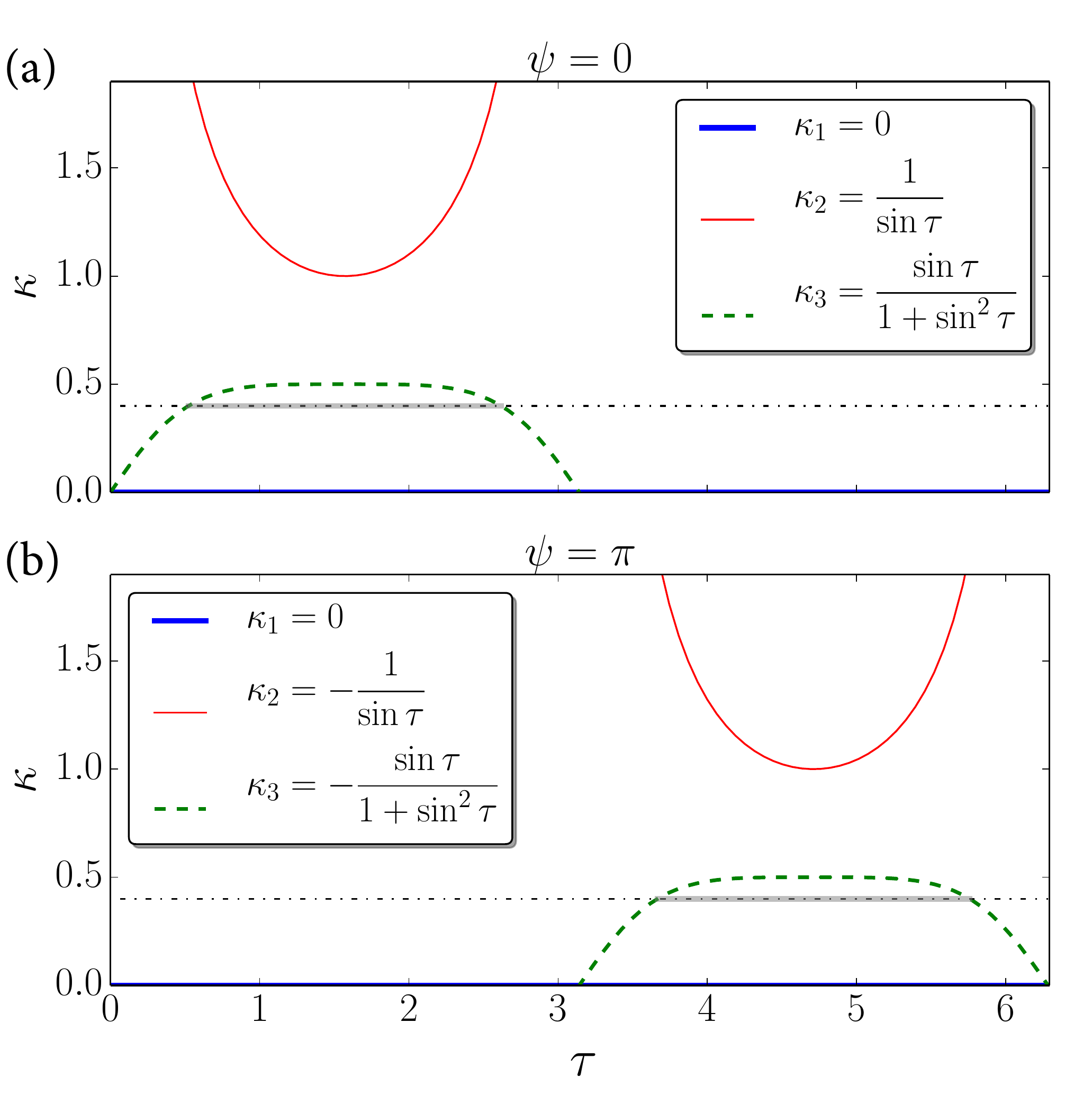}
\caption{\label{bif_diagram}\raggedright The saddle-node bifurcation curves ($\lambda_1=0$) of the characteristic equation\,\eqref{deteq}for $\Delta=0$ in the plane of coupling strength $\kappa$ and delay time $\tau$ given by Eq.\,\eqref{bif_in} for the in-phase mode (a) and Eq.\,\eqref{bif_anti} for the anti-phase mode (b). The horizontal dashed-dotted line represents $\kappa=0.4$ and its grey shaded part is the unstable region with $\lambda_i>0$. The other parameters are $\omega_1 = \omega_2=1$, $\mu = 0.1$, $\gamma=1$.}
\end{figure}

{\em In- and anti-phase mode -}
Eqs.\,(\ref{R1}) and (\ref{Psi}) possess two equilibrium solutions, an in-phase and an anti-phase mode, as we will demonstrate below. As we recognize from Fig.\,\ref{goal} the center of the synchronization region plays a special role. This motivates a first investigation of the solutions and their stability for vanishing detuning $\Delta=0$. Such a special parameter setting reduces the technical difficulties and nevertheless allows us to make a qualitative and qualitative analysis of our problem. We find the following equilibrium points for Eqs.\,(\ref{R1}) and (\ref{Psi}) which in turn correspond to the in-phase and anti-phase mode of Eq.\,\eqref{ausgang}:
\begin{equation}
\label{loes1}
\begin{split}
\psi_{in}=0 \Leftrightarrow R_1= R_2=\frac{2}{\sqrt{\gamma}}\sqrt{1-{\kappa}\sin \tau},\\
\psi_{anti}=\pi \Leftrightarrow R_1= R_2=\frac{2}{\sqrt{\gamma}}\sqrt{1+{\kappa}\sin \tau}.
\end{split}
\end{equation}
In order to determine the stability of the in-phase and anti-phase mode, we linearize Eqs.\,(\ref{R1}),(\ref{Psi}) around the equilibrium points, which gives the Jacobian matrix $J$ of the system\,(\ref{R1}),(\ref{Psi}):
\begin{widetext}
\begin{equation}
\label{Jey}
J=\left(
\begin{array}{ccc}
 1-\frac{3}{4}  {\gamma} R_1^2 & -\kappa \sin(\psi+\tau) & -\kappa R_2 \cos(\psi+\tau) \\
 \kappa \sin(\psi-\tau) & 1-\frac{3}{4}  {\gamma} R_2^2 & \kappa R_1 \cos(\psi-\tau) \\
 \kappa \left[\frac{R_2 \cos(\psi+\tau)}{R_1^2}+\frac{\cos(\psi-\tau)}{R_2}\right] & -\kappa \left[\frac{\cos(\psi+\tau)}{R_1}+\frac{R_1 \cos(\psi-\tau)}{R_2^2}\right] & \kappa \left[\frac{R_2 \sin(\psi+\tau)}{R_1}-\frac{ R_1 \sin(\psi-\tau)}{R_2}\right]\\
\end{array}
\right).
\end{equation}
\end{widetext}
The eigenvalues $\lambda_{i}$, $i=1,2,3$ of the Jacobian matrix evaluated at these equilibrium points determine their linear stability. They are calculated from the characteristic equation
\begin{eqnarray}
\label{deteq}
\det(J-\lambda_{i}\mathsf{I}) = 0.
\end{eqnarray}
in dependence on the system parameters $\gamma$, $\kappa$, and $\tau$. Let us first
consider the in-phase mode $\psi_{in}=0$ in Eq.\,\eqref{loes1}:
\begin{equation}
\label{chareq}
\begin{split}
(\lambda+2-2\kappa \sin \tau) [6\kappa^2+ \lambda(\lambda+2)-\\
2\kappa \left( \kappa \cos 2\tau + (3 \lambda + 2)\sin \tau \right)  ]= 0.
\end{split}
\end{equation} 
This equation depends upon the two parameters $\kappa$ and $\tau$. The boundary of stability with respect to saddle-node bifurcations is given by the condition $\lambda_1=0$, which defines the generic saddle-node bifurcation curves in the ($\kappa$, $\tau$) plane:
\begin{eqnarray}
\label{4kappa}
\kappa \left[\kappa-\sin \tau\left(1-2\kappa \sin \tau + \kappa^2(1+\sin^2 \tau)\right)\right]=0.
\end{eqnarray}
We obtain three solution branches for $\kappa$ fulfilling Eq.\,\eqref{4kappa}:
\begin{eqnarray}
\label{bif_in}
\kappa_1 &=&0,\nonumber\\
\kappa_2&=&\frac{1}{\sin \tau},\\
\kappa_3&=&\frac{\sin \tau}{1+\sin^2 \tau}.\nonumber
\end{eqnarray}
In the case of the anti-phase mode $\psi_{in}=\pi$ in Eq.\,\eqref{loes1} the characteristic equation\,\eqref{deteq} reads:
\begin{equation}
\label{chareq_anti}
\begin{split}
(\lambda+2+2\kappa \sin \tau) [6\kappa^2+ \lambda(\lambda+2)-\\
2\kappa \left( \kappa \cos 2\tau - (3 \lambda + 2)\sin \tau \right)  ]= 0.
\end{split}
\end{equation} 
The generic saddle-node bifurcation curves in the ($\kappa$, $\tau$) plane is given by:
\begin{eqnarray}
\label{bif_anti}
\kappa_1 &=&0,\nonumber\\
\kappa_2&=&-\frac{1}{\sin \tau},\\
\kappa_3&=&-\frac{\sin \tau}{1+\sin^2 \tau}.\nonumber
\end{eqnarray}
Note that $\kappa_1=0$ represents an uncoupled system. The bifurcation curves $\kappa_{2/3}$ in Eqs.(\ref{bif_in}) and (\ref{bif_anti}) separate the regions of stable and unstable equilibrium in the ($\kappa$, $\tau$) plane for in-phase and anti-phase mode, respectively; they are represented in Fig.\,\ref{bif_diagram}(a), (b), respectively. By fixing a value of $\kappa$, e.g., $\kappa=0.4$ (horizontal dash-dotted line), one can trace the change of stability as $\tau$ is changed. For $\tau=0$ Eqs.\,(\ref{chareq}), (\ref{chareq_anti}) reduce to
\begin{eqnarray}
\label{tau_0}
\lambda^3+4 \lambda^2+4 \lambda (\kappa^2+1)+8 \kappa^2=0
\end{eqnarray}
which has no solution $\lambda$ with positive real part, hence both equilibria are stable for $\tau=0$.
The horizontal dash-dotted line $\kappa=0.4$ intersects with the bifurcation curve $\kappa_3=\pm\frac{\sin \tau}{1+\sin^2 \tau}$ as shown in Fig.\,\ref{bif_diagram}(a), and hence for $\frac{1}{6}\pi < \tau < \frac{5}{6}\pi$ the in-phase mode becomes unstable ($\operatorname{Re}\lambda_i>0$), whereas the anti-phase mode becomes unstable for $\frac{7}{6}\pi < \tau < \frac{11}{6}\pi$, see Fig.\,\ref{bif_diagram}(b). In the remaining ranges of $\tau$ the in-phase and anti-phase modes, respectively, are stable. Note that bistability of in-phase and anti-phase mode occur around $\tau=0$ and $\tau=\pi$, as also visible in Fig.\,\ref{goal} for $\tau=0.1 \pi$.
In Fig.\,\ref{adlerplot2}(a) a smaller value $\kappa=0.2$ is chosen, and hence for $\tau=0.1 \pi$ the in-phase mode is unstable and no bistability exists, in full agreement with Fig.\,\ref{bif_diagram}.

{\em Synchronization frequency -} 
The describing function method yields Eq.\,\eqref{freql} which determines the synchronization frequency $\omega_s$, if the phase difference $\psi$ is known. The method of averaging yields the generalized Adler equation\,\eqref{adler} determining the dynamics of $\psi$. In Fig.\,\ref{adlerplot2}a we have found numerically with the help of the Adler equation a stable ($\psi=\pi$) and an unstable ($\psi=0$) equilibrium point of $\psi$.
\begin{figure}
\includegraphics[width=1.05\linewidth]{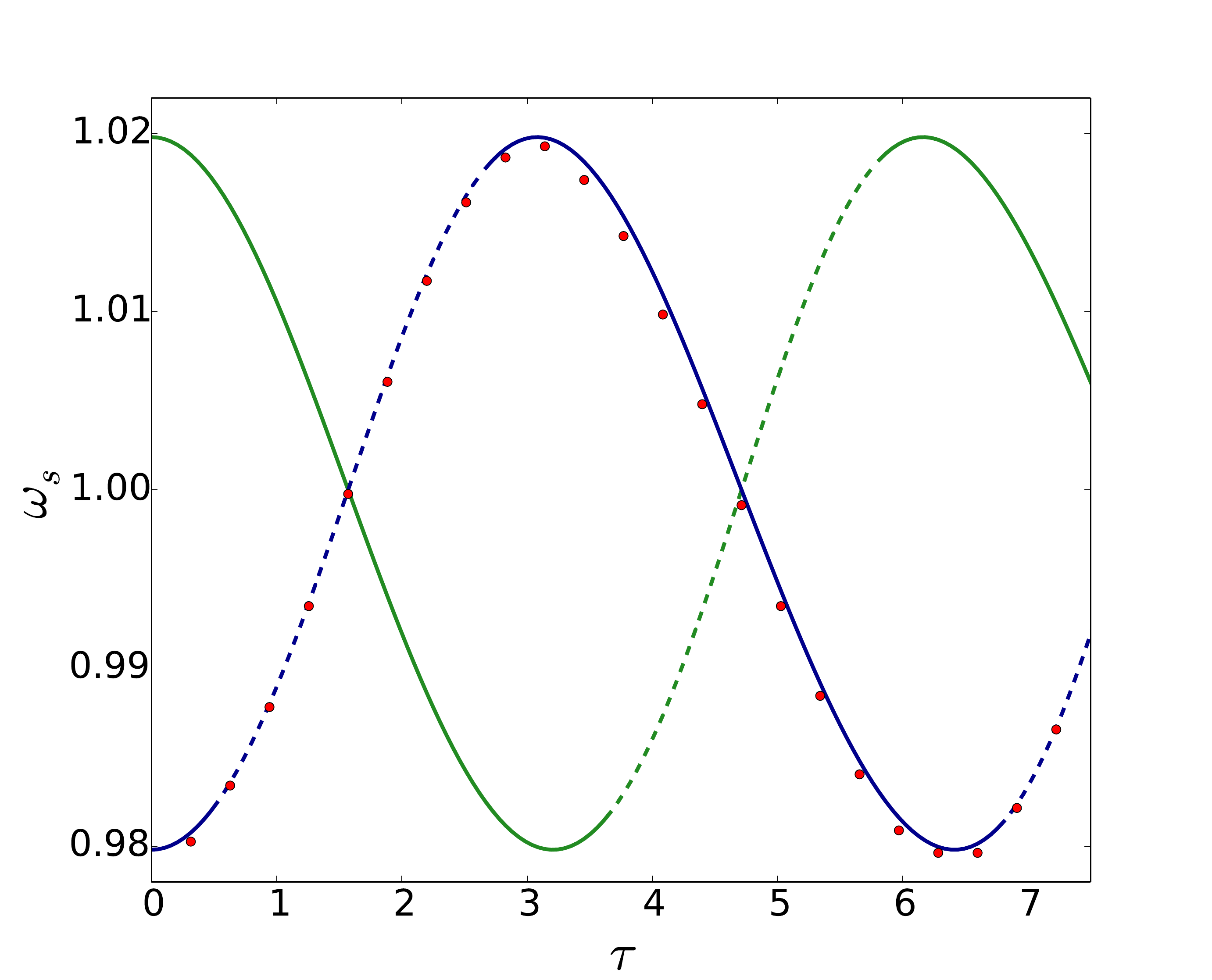}
\caption{\label{comp1}\raggedright Comparison of the analytic (line) and numeric (dots) results of the synchronization frequency $\omega_s$ versus the delay time $\tau$ for $\Delta=0, \omega_1=\omega_2=1$, $\mu = 0.1$, $\gamma=1$, $\kappa=0.4$. The analytic solution gives an in-phase (dark blue line) and an anti-phase mode (light green line), while the numeric solution (dots) with symmetric initial conditions only reproduces the in-phase mode. Solid line means stable solution, whereas dashed line stands for an unstable one as shown in Fig.\,\ref{bif_diagram}.}
\end{figure}
In order to compare the numerical simulations with the results of the describing function method, we set $\omega_1=\omega_2=1$, i.e., 
$\Delta=0$, in which case the equilibrium solutions of $\psi$ are given by Eq.\,\eqref{loes1}, and hence Eq.\,\eqref{freql} can be simplified to 
\begin{equation}
\label{omega_simpl}
\begin{split}
(\omega_s^2-1)^2= \mu^2\kappa^2\cos^2(\omega_s\tau)
\end{split}
\end{equation}
or 
\begin{equation}
\label{omega_simpl2}
\begin{split}
\omega_s^2=1 \pm\mu\kappa\cos(\omega_s\tau)
\end{split}
\end{equation}
where $+$ and $-$ correspond to anti-phase and in-phase oscillations, respectively. In Fig.\,\ref{comp1} we plot the synchronization frequency $\omega_s$ versus the time delay $\tau$ for $\Delta=0$. The congruence between the numerical (from Eq.\,(\ref{ausgang})) and analytical result for the in-phase mode (dark blue line, from Eq.\,(\ref{omega_simpl2})) is excellent. It is remarkable that the synchronization frequency $\omega_s$ is modulated around the single oscillator frequencies $\omega_i=1$ in dependence upon the delay time. For small delay time, for instance, the in-phase oscillation frequency is lowered, while the anti-phase oscillation frequency (light green line) is increased. The stability of the two branches changes as $\tau$ is varied, as discussed above, see Fig.\,\ref{bif_diagram}: At the extrema of the frequency curve in Fig.\,\ref{comp1} we can find bistability.
 Note that for symmetric initial conditions the in-phase mode is found as numerical solution for all delay times although it is unstable in part of the $\tau$ range (but there for any non-symmetric initial conditions the anti-phase mode would be found). In Fig.\,\ref{goal} the upper frequency branch in the synchronization region stays in the stable anti-phase mode for non-zero detuning $\Delta$ (in congruence with experimental data \cite{ABE06}), whereas the lower branch, i.e., the stable in-phase mode, which is close to its instability point, is only observable in a small range of $\Delta$.

%%%%%%%%%%%%%%%%%%%%%%%%%%%%%%%%%%%%%%%%%%%%%%%%%%%%%%%%%%%%%%%%%%%%%
\section{Comparison with acoustic experiments} \label{sec:comparison}
\begin{figure}
\includegraphics[width=1.0\linewidth]{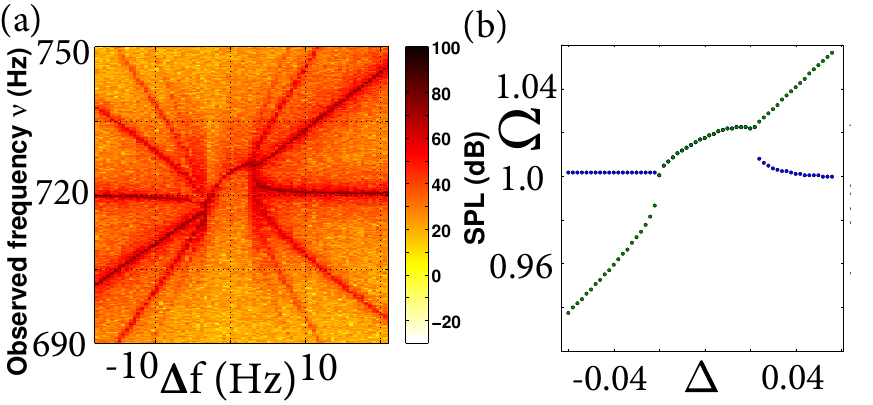}
\caption{\label{realpipeex}\raggedright Comparison of synchronization region in experiment and theory: (a) experimentally observed sound pressure level (SPL) in the plane of observed frequency $\nu$ vs. frequency detuning $\Delta f$ (in Hz) \cite{FIS14} and (b) numerically calculated angular frequency $\Omega$ vs. dimensionless detuning $\Delta$ for $\omega_2=1$, $\mu = 0.1$, $\gamma=1$, $\kappa=0.4$, $\tau=1.1\pi$.}
\end{figure}

\begin{figure}
\includegraphics[width=1.0\linewidth]{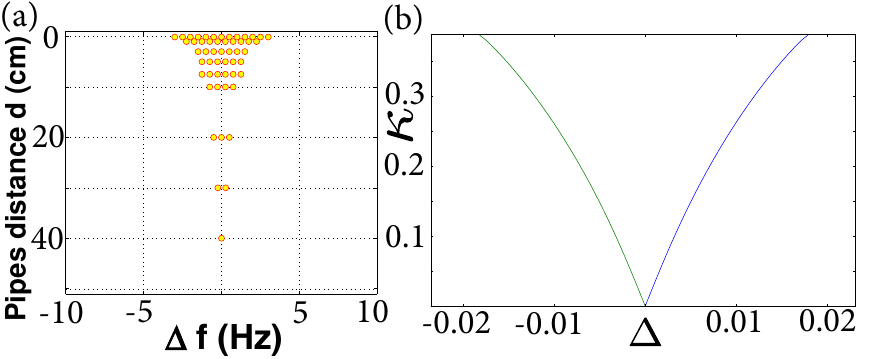}
\caption{\label{realanal}\raggedright Comparison of experiment and theory: Arnold tongue in the plane of coupling strength $\kappa$ vs. detuning $\Delta$: (a) experiment \cite{FIS14}, where the coupling strength is given by the distance $d$ of the organ pipes, and (b) analytic result for $\omega_2=1$, $\mu = 0.1$, $\gamma=1$, and delay time $\tau=0.1\pi$.}
\end{figure}

A comparison of a complex experiment with a simple oscillator model is an ambitious endeavor: On one hand there is an organ pipe with a whole spectrum of overtones and a complicated aeroacoustic behavior, on the other hand we consider a simple Van der Pol oscillator. Nevertheless, such a simple model can already exhibit complicated dynamical scenarios, as demonstrated above. Qualitatively, these scenarios agree well with the experiment, as we show by comparing the graphs below. Consequently, our model provides a profound comprehension of the dynamical behavior observed in organ pipes. This supports our point of view that complex behavior can emerge from simple, low-dimensional systems, as it is generally accepted in dynamical systems theory.\\
For visual comparison of experiment and theory, we show the synchronization region versus the detuning frequency of the two oscillators in Fig.\,\ref{realpipeex}. Both plots show very similar features. Especially the behavior of the transition regions at the two boundaries of the locking interval is remarkable, as well as the concave curvature of the synchronization region itself. 
In Fig.\,\ref{realanal} we compare the experimentally observed and the analytically calculated Arnold tongue in the plane of the coupling strength $\kappa$ and the detuning $\Delta$. Experimentally, the coupling strength is determined by the distance of the two organ pipes. The analytical calculation proceeds as described in Sect. II. As a result we obtain an Arnold tongue with nonlinear, curved boundaries, see Fig.\,\ref{realanal}b. This is a remarkable result which occurs already for a small delay time $\tau$ and coincides well with experiments \cite{FIS14,FIS16}, see Fig.\,\ref{realanal}a.  The curvature of the boundaries may be further adjusted in the calculations by replacing the constant $\kappa$ by a $\tau$-dependent coupling strength $\kappa(\tau)$, see Appendix.

%%%%%%%%%%%%%%%%%%%%%%%%%%%%%%%%%%%%%%%%%%%%%%%%%%%%%%%%%%%%%%%%%%%

\section{Conclusion} \label{sec:conclusion}
In this paper we have investigated the synchronization of organ pipes 
using the tools of nonlinear dynamics. Particular attention has 
been paid to the delay in the coupling which naturally occurs due to a finite 
distance of pipes. We have used a simplified nonlinear oscillator model for the pipes,
i.e., two coupled Van der Pol oscillators which interact by a dissipative, direct, delayed 
coupling. To understand theoretically the dynamics of the system, we 
have analyzed the locking scenarios. Further, we have numerically integrated the system, and have found that
the solution agrees well with existing experiments. On this basis, we have systematically 
varied the coupling parameters, namely the coupling strength $\kappa$ 
and the coupling delay $\tau$.\\
For a deeper understanding of the various bifurcation scenarios we 
have developed and extended two complementary analytical approaches:
By the method of averaging we obtain a generalized Adler equation for the phase dynamics, which allows us 
to study the stability of the equilibria corresponding to frequency 
locking of the oscillators.  However, the averaging 
method does not provide information about the frequency in 
the locking region. The frequency, in contrast, can be found by the 
describing function method which allows us to determine the 
synchronization frequency and hence explain the curvature of the 
frequency vs.\,detuning which is found in the numerical simulations as well as in the 
experiments (see Fig.\,\ref{realpipeex}). Altogether these approximations 
 provide a detailed and complete analytic picture of both relative phase and frequency. 
In general we obtain excellent agreement of our analytic results 
with the numerical simulations and with experiments.\\
A detailed bifurcation analysis has affirmed the existence of in- and 
anti-phase synchronization. In each case the synchronization frequency 
has a different value which is in perfect accordance with our analytic 
calculations. The behavior of the boundaries of the Arnold tongue in the 
plane of coupling strength $\kappa$ and detuning $\Delta$ depends on the interplay of the 
coupling strength and the coupling delay time $\kappa$ . In general, the nonlinear 
interdependence of $\kappa$ and $\tau$ leads to curved boundaries in the ($\kappa$, $\Delta$) plane, 
which is also clearly confirmed by experimental data. 

It is interesting to note that there is some similarity of our delayed coupling with the viscoelastic coupling, which has been used in a recent study of two modified Van der Pol oscillators with the aim to describe cardiac synchronization~\cite{STE17}. This viscoelastic coupling is modeled within the Maxwell model of viscous creep by an additional differential equation describing a harmonic spring in series with a linear damper of damping rate (viscosity) $\gamma$. This linear inhomogeneous differential equation can be eliminated using 
a Green's function approach, thereby introducing a distributed delay in the coupling of the two oscillators with an exponential delay kernel with decay rate $\gamma$ corresponding to a temporal memory ~\cite{KYR11,KYR13,KYR14}. In this viscoelastic model also in-phase and anti-phase synchronization scenarios were found.

\begin{acknowledgments}
This work was partially supported by Deutsche Forschungsgemeinschaft in the
framework of SFB 910. We are grateful to Jost Fischer and Natalia Spitha for fruitful discussions.
\end{acknowledgments}

\section*{Appendix} \label{sec:appendix}

\begin{figure}
\includegraphics[width=1.02\linewidth]{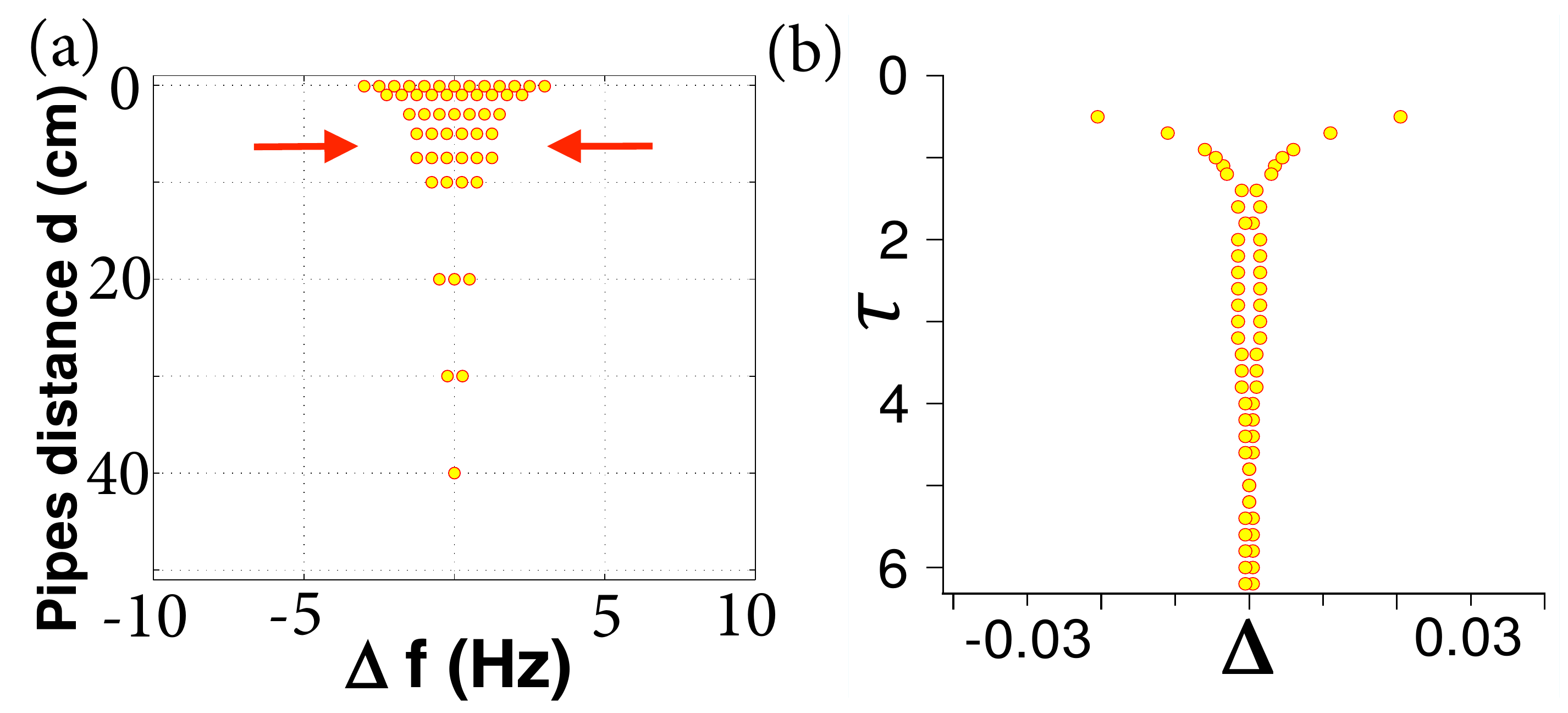}
\caption{\label{coupling_function}\raggedright Comparison of experiment and theory: Arnold tongue in the plane of coupling strength $\kappa$ vs. detuning $\Delta$: (a) experiment \cite{FIS14}, where the coupling strength is given by the distance $d$ of the organ pipes, and (b) numerical result of Eq.\,\eqref{ausgang_end} with $\omega_2=1$, $\mu = 0.1$, $\gamma=1$, $\kappa_1=\kappa_2=0.04$. The red arrows in (a) indicate the non-monotonic behavior of the Arnold tongue.}
\end{figure}

For a more refined modeling of the non-monotonic behavior of the Arnold tongue as observed in experiment, see Fig.\,\ref{coupling_function}a for $d=5cm$, instead of a constant coupling factor $\kappa$ as in Eq.\,\eqref{ausgang}, a coupling strength $\kappa(\tau)$ which depends on the delay time $\tau$, should be used. The coupling is delayed, because the sound travels a certain distance $d$ between the pipes. The coupling strength depends on that distance, since the sound wave is attenuated according to the radiation of a spherical wave emitted from the pipe mouth. Within the coupling strength $\kappa(\tau)$ we have a near-field term ($\propto\frac{1}{{\tau}^2}$) and a far-field term ($\propto\frac{1}{{\tau}}$) with coefficients $\kappa_1, \kappa_2 > 0$:
\begin{equation}
\label{coupling_delay}
\kappa(\tau)= \frac{\kappa_1}{{\tau}^2}+\frac{\kappa_2}{{\tau}}.
\end{equation}
By replacing $\kappa$ in Eq.\,\eqref{ausgang} by Eq.\,\eqref{coupling_delay} 
\begin{equation}
\label{ausgang_end}
\ddot{x}_i+{\omega_i}^2x_i-{\mu}\left[\dot{x}_i-\dot{f}(x_i)+{\kappa}(\tau)x_j(t-\tau)\right]=0,
\end{equation}
we are able to model the boundaries of the Arnold tongue more realistically (see Fig.\,\ref{coupling_function}b).

%\bibliography{ref}
%\bibliographystyle{prwithtitle}
%\bibliographystyle{prsty-fullauthor}

\end{document}